\documentclass[twocolumn,12pts,aps,showpacs]{revtex4}
\usepackage{amsmath}

\begin{document}
\title{Charged perfect fluid disks as sources of Taub-NUT-type spacetimes}

\author{Gonzalo Garc\'\i a-Reyes}\email[Email address: ]{ggr1970@yahoo.com}
\author{Guillermo A. Gonz\'{a}lez}\email[Email address: ]{guillego@uis.edu.co}

\affiliation{Escuela de F\'{\i}sica, Universidad Industrial de Santander, A. A.
 678, Bucaramanga, Colombia}


\begin{abstract}

The interpretation of a family of electrovacuum stationary Taub-NUT-type 
fields in terms of finite charged perfect fluid disks is presented. The
interpretation is mades by means of an ``inverse problem" approach used to
obtain disk sources of known solutions of the Einstein or Einstein-Maxwell
equations.  The diagonalization of the energy-momentum tensor of the disks is
facilitated in this case by the fact that  it  can be written as an upper right
triangular matrix.  We find that the  inclusion of electromagnetic fields
changes significatively the different material properties of the  disks and so
we can obtain, for some values of the parameters, finite  charged perfect fluid
disks that are in agreement with all the energy conditions.

\end{abstract}

\maketitle

\section{INTRODUCTION}

Several methods are known to exactly solve the Einstein and  Einstein-Maxwell
field equations, or to generate new exact  solutions from simple known
solutions \cite{KSHM}. However, the above mentioned methods in general lead to 
solutions without a clear physical interpretation or to  solutions that depend
of many parameters without a clear physical meaning. Accordingly, it is of
importance to have  some appropriate procedures to obtain physical
interpretations of these exact solutions. So, in the past years such  
procedures have been developed for static and stationary  axially symmetric
solutions in terms of thin and, more  recently, thick disk models.

Stationary or static axially symmetric exact solutions of  Einstein  equations
describing relativistic thin disks are of  great astrophysical importance since
they can be used as  models of certain stars, galaxies and accretion disks.
These were first studied by Bonnor and Sackfield \cite{BS},  obtaining
pressureless static disks, and then by Morgan and  Morgan, obtaining static
disks with and without radial  pressure \cite{MM1,MM2}. In connection with
gravitational collapse, disks were first studied by Chamorro, Gregory, and 
Stewart \cite{CHGS}. Disks with radial tension have been also  studied
\cite{GL1}. Recently, more realistic models of thin disks and thin disks with
halos  made of perfect fluids were considered in \cite{VL1}. 

Several classes of exact solutions of the  Einstein field equations
corresponding to static and  stationary thin disks have been obtained by
different authors  \cite{LP,LO,LEM,LL1,BLK,BLP,BL,LL2,LL3,KLE,GL2}, with or 
without radial pressure. Except for the pressureless disks, generally the other
disks have as source matter with azimuthal pressure (tension) different from
the radial pressure (tension).  However, in some cases these disks can be
interpreted as the superposition of two counterrotating perfect fluids.  A
detailed study of the counterrotating model for the case of static thin disks 
is presented in \cite{GE}. 

Disk sources for stationary axially symmetric spacetimes with  magnetic fields
are also of astrophysical importance mainly in the study of neutron stars,
white dwarfs and galaxy formation. Although disks with electric fields do not
have clear  astrophysical importance, their study may be of interest in the
context of exact solutions. Thin disks  have been discussed as sources for
Kerr-Newman fields \cite{LBZ}, magnetostatic  axisymmetric fields \cite{LET1},
conformastationary metrics  \cite{KBL}, and electrovacuum static
counterrotating disks \cite{GG}.

In all the above cases, the disks are obtained by an ``inverse problem''
approach, called by Synge the ``{\it g-method}'' \cite{SYN}. The method works
as follows: a solution of the vacuum Einstein equations is taken, such that
there is a discontinuity in the derivatives of the metric tensor on the plane
of the disk, and the energy-momentum tensor is obtained from the Einstein 
equations. The physical properties of the matter distribution are then studied
by an analysis of the surface energy-momentum tensor so obtained. Very recently
the above procedure has been generalized in order to obtain thick disk models
\cite{GL3} and also non-axisymmetric planar distributions of charged dust
\cite{VL2}. As we can see, by this ``inverse problem'' approach we can do
physical interpretations of many known solutions of the  Einstein and
Einstein-Maxwell vacuum equations, in the sense that the thin disks can act as
exact sources for the space-time metrics given by the vacuum solutions.

The purpose of this paper is the interpretation of a family of electrovacuum
stationary Taub-NUT-type spacetimes in terms of finite charged perfect fluid
disks. The paper is structured as follows. We first present, in Sec. II, the
family of stationary Taub-NUT-type solutions of the Einstein-Maxwell equations
that we will consider. Then, in Sec. III, we present a summary of the procedure
to  construct  models of  disks with nonzero radial pressure. We obtain
expressions for  the surface energy-momentum  tensor and the surface current
density of the disks. In particular we find explicit expressions for the
surface energy  density, the pressure and the surface charge electric density.
Finally, the disks are interpreted in terms of a charged perfect fluid. In Sec.
IV we analyze the physical properties of the disks and we show that, for some
values of the parameters, the disks are in agreement with all the energy
conditions. We also study the motion of the disks and their stability against
radial perturbations is analyzed. Finally, in Sec. V,  we summarize our main
results.   

\section{Einstein-Maxwell equations and Taub-NUT-type fields}

The line element of a stationary axisymmetric field can be written in
quasi-cylindrical coordinates (t,$\varphi,r,z$) in the form  
\begin{equation}
ds^2 = - e^{2 \Psi} (dt + {\cal W} d\varphi)^2 \ + \ e^{- 2 \Psi}[ R^2 
d\varphi^2 + e^{2 \Lambda} (dr^2 + dz^2)], \label{eq:met}
\end{equation}
where $R$, $\Psi$, ${\cal W}$, and $\Lambda$ are functions of  $r$ and $z$
only. The vacuum Einstein-Maxwell equations,  in  geometrized units in which $8
\pi G = c = \mu _0 =  \varepsilon _0 = 1$,  are given by 
\begin{subequations}\begin{eqnarray}
&   &   G_{ab} \  =  \ T_{ab},  \\
&   &     \nonumber       \\
&   &    F^{ab}_{ \ \ \ ; b} \ =  \ 0,     
\end{eqnarray}\label{eq:einmax}\end{subequations}
with
\begin{subequations}\begin{eqnarray}
T_{ab} \  &=&  \ F_{ac}F_b^{ \ c} - \frac{1}{4} g_{ab} F_{cd}F^{cd}, 
\label{eq:et} \\
&   &     \nonumber    \\
F_{ab} \ &=&  \  A_{b,a} -  A_{a,b}.
\end{eqnarray}\end{subequations}

For the metric (\ref{eq:met}), the  Einstein-Maxwell equations in vacuum imply
that  $R$ satisfies  the Laplace's  equation
\begin{equation}
R_{,rr} \ + \ R_{,zz} \ = \ 0,  \label{eq:lap}
\end{equation}
so that the function $R$ can be considered as the real part of an analytical  
function  $F (\nu) = R(r,z) + i Z(r,z)$, where $\nu = r + iz$. Thus the
function  $F (\nu)$ defines  a  conformal transformation
\begin{equation}\begin{array}{ccc}
r & \rightarrow & R (r, z) , \\
&	&	\\
z & \rightarrow & Z (r, z) , 
\end{array}\label{eq:conf}\end{equation}
in such a way that the metric   (\ref{eq:met}) takes the usual
Weyl-Lewis-Papapetrou form
\begin{equation}
ds^2 = - \ e^{2 \tilde \Psi} (dt + \tilde {\cal W} d\varphi)^2 \ + \ e^{- 2
\tilde \Psi}[ R^2  d\varphi^2 + e^{2 \tilde \Lambda} (dR^2 + dZ^2)],
\label{eq:newmet}
\end{equation}
where  $R$ and $Z$  are the Weyl's  canonical coordinates. In  this
coordinates,  the field equations (\ref{eq:einmax}) are  equivalent to the
usual complex  Ernst equations \cite{E2}
\begin{subequations}\begin{eqnarray}
f \Delta {\cal E} &=& (\nabla {\cal E} + 2\Phi^\ast\nabla\Phi) \cdot
\nabla{\cal E}, \\
&  & \nonumber  \\
f \Delta \Phi &=& (\nabla {\cal E} + 2 \Phi^\ast \nabla \Phi) \cdot \nabla
 \Phi, 
\end{eqnarray}\label{eq:ernst}\end{subequations}
where  $\Delta$ and  $\nabla$ are  the  standard differential  operators in
cylindrical coordinates and $f= e^{2 \tilde \Psi} $. The metric functions and
the electromagnetic  potentials are obtained from the relations
\begin{subequations}\begin{eqnarray}
f &=& \mathrm{Re} \ {\cal E}  +\Phi \Phi^* ,   \\
&&	\nonumber	\\
4 f^2 \tilde \Lambda _{,\zeta} &=& \sqrt{2} R ( {\cal E}_{,\zeta} + 2 \Phi ^*
\Phi_{,\zeta} ) ({\cal E}_ {,\zeta}^* + 2 \Phi \Phi^*_{,\zeta} ) - 4f
\Phi_{,\zeta} \Phi^*_ {,\zeta},         \\
&&	\nonumber	\\      
f^2 \tilde {\cal W} _{,\zeta} &=& R [ i({\rm Im}{ \cal E})_{,\zeta} + \Phi^*
\Phi_{,\zeta} - \Phi \Phi^*_{,\zeta} ], \\
&&	\nonumber	\\      
\tilde A_t &=& \sqrt 2 \ {\rm Re} \ \Phi ,\\
&&	\nonumber	\\      
f \tilde  A_{\varphi,\zeta} &=&  \sqrt 2 [ i R ( {\rm Im}  \Phi )_{,\zeta} +
\tilde {\cal W} \tilde A_{t,\zeta} ] ,
\end{eqnarray}\label{eq:lambda}\end{subequations}
where $\sqrt{2} \zeta = R + i Z$, so that $\sqrt{2} \partial_{,\zeta} =
\partial_{,R} - i \partial_{,Z}$.

We will consider a simple stationary Taub-NUT-type solution to the above system
of equations given by 
\begin{subequations}\begin{eqnarray}
\tilde \Psi \ &=& \ \frac 12 \ln \left[ \frac{x^2 -1}{x^2 +  2ax + 1+ c^2}
\right],  \\
& &   \nonumber      \\
\tilde \Lambda \ & = & \ \frac 12 \ln \left [ \frac {x^2 -1}
{x^2-y^2} \right],  \\ 
& &   \nonumber      \\ 
\tilde {\cal W} \ &=& \  2kby,     \\
& &   \nonumber      \\                     
\tilde A_{t} \ &=& \  \frac {\sqrt 2 c} {\sqrt{1+c^2}} \left 
[\frac { 1 + c^2 + ax }{ x^2 +2ax +1 + c^2} \right ],  \\
& &   \nonumber      \\
\tilde A _{\varphi} \ & = & \  \frac {\sqrt 2 c kby}{\sqrt{1+c^2}} \left
[\frac { 1 + c^2 - x^2 }{ x^2 +2ax +1 + c^2} \right ],
\end{eqnarray}\label{eq:tnt}\end{subequations}
where $a^2+b^2 = 1+c^2$, being $c$ the parameter that controls the
electromagmetic field. The prolate  spheroidal coordinates, $x$ and $y$, are
related to the Weyl coordinates by 
\begin{equation}
R^2   = k^2 (x^2-1)(1-y^2),  \quad  \quad Z  = kxy, 
\label{eq:coorp}
\end{equation}
where  $1 \leq x \leq \infty$ and $0  \leq y \leq 1 $.  This solution can be
generated, in these coordinates,  using  the well-known  complex potential
formalism proposed by Ernst \cite{E2} from the   Taub-NUT vacuum solution
\cite{KSHM}.  Note that  when $c=0$ this solution reduces to the Taub-NUT
vacuum solution. 

Once the above solution is known, we can obtain a solution of the field
equations (\ref{eq:einmax}) in the original coordinates  by setting 
\begin{subequations}\begin{eqnarray}
 R(r,z) &=&   {\rm Re} \ F(\nu) ,     \\
&&	\nonumber	\\
\Psi (r,z) \ &=& \ \tilde \Psi (R,Z) , \label{eq:com3} \\
&&	\nonumber	\\
\Lambda (r,z) \ &=& \ \tilde \Lambda (R,Z) \ + \ \ln
|F' (\nu)| , \label{eq:com4} \\
&&	\nonumber	\\
{\cal W} (r,z) \ &=& \ \tilde {\cal W} (R,Z) ,   \\
&&	\nonumber	\\
A_t (r,z) \ &=& \ \tilde A_t (R,Z),  \\
&&	\nonumber  	\\
A_\varphi (r,z) \ &=& \ \tilde A_\varphi (R,Z),
\end{eqnarray}\label{eq:comf}\end{subequations}
where  $F'=dF/d\nu$. Note also that the solution to the original
Einstein-Maxwell equations (\ref{eq:einmax}) is not completely determined as we
do not take any especific choice for $F ( \nu)$. We will do this in the next
section. 

\section{A family of finite charged perfect fluid disks}

In order to obtain a solution of (\ref{eq:einmax})  
representing a thin disk at $z=0$, we assume that the  
components of the metric tensor and the electromagnetic 
potentials are continuous across the disk, but with first 
derivatives discontinuous on the  plane $z=0$, with 
discontinuity functions
\begin{eqnarray}
b_{ab} \ &=& g_{ab,z}|_{_{z = 0^+}} \ - \ g_{ab,z}|_{_{z = 0^-}} \ = \ 2 \
g_{ab,z}|_{_{z = 0^+}}  ,  \nonumber                   \\
 &&         \nonumber                   \\
a_a \ &=& A_{a,z}|_{_{z = 0^+}} \ - \ A_{a,z}|_{_{z = 0^-}} 
\ = \ 2 \ A_{a,z}|_{_{z = 0^+}}.   \nonumber                  
\end{eqnarray}
Thus, the Einstein-Maxwell equations yield an energy-momentum tensor $T_{ab}=
T^{\mathrm{elm}}_{ab} +  T^{\mathrm {mat}}_{ ab}$,  where $T^{\mathrm
{mat}}_{ab} = Q_{ab} \ \delta (z)$,   and a current density $J_a = j_a  \delta
(z)= - 2 e^{2 (\Psi - \Lambda)} A_{a,z} \delta (z)$, where  $\delta (z)$ is
the  usual Dirac function with support on the disk.  $T^{\mathrm { elm}}_{ab}$
is the electromagnetic tensor defined in Eq.  (\ref{eq:et}), $j_a$ is the
current density on the plane  $z=0$, and 
$$
Q^a_b = \frac{1}{2}\{b^{az}\delta^z_b - b^{zz}\delta^a_b + 
g^{az}b^z_b -
g^{zz}b^a_b + b^c_c (g^{zz}\delta^a_b - g^{az}\delta^z_b)\}
$$
is the distributional energy-momentum tensor. The ``true'' surface
energy-momentum tensor (SEMT) of the  disk, $S_{ab}$, and the ``true'' surface
current density, $\mbox{\sl j}_a$, can be obtained through the relations
\begin{subequations}\begin{eqnarray}
S_{ab} &=& \int T^{\mathrm {mat}}_{ab} \ ds_n \ = \ e^{ \Lambda - \Psi} \
Q_{ab} \ ,   \\
\mbox{\sl j}_a  &=& \int J_{a}  \ ds_n \ = \ e^{ \Lambda - \Psi} \ j_a \ , 
\end{eqnarray}\end{subequations}
where $ds_n = \sqrt{g_{zz}} \ dz$ is the ``physical measure'' of length in the
direction normal to the disk.

For the metric (\ref{eq:met}), the nonzero components of $S_a^b$ are
\begin{subequations}\begin{eqnarray}
&S^0_0 &= \ \frac{e^{\Psi - \Lambda}}{ R^2} \left [ 2 R^2
(\Lambda,_z - \ 2 \Psi,_z) + \ 2 RR,_z - \ e^{4\Psi}
{\cal W} {\cal W},_z \right ] , \label{eq:emt1}  \\
	&	&	\nonumber	\\
&S^0_1 &= \ \frac{e^{\Psi - \Lambda}}{ R^2} \left [ 2 R 
{\cal W}
(R,_z - \ 2 R \Psi,_z) - \ ( R^2 + \ {\cal W}^2
e^{4\Psi} ) {\cal W},_z \right ] , \label{eq:emt2}  \\
	&	&	\nonumber	\\
&S^1_0 &= \ \frac{e^{\Psi - \Lambda}}{ R^2} \left[ e^{4\Psi} 
{\cal W},_z \right ], \label{eq:emt3} \\
	&	&	\nonumber	\\
&S^1_1 &= \ \frac{e^{\Psi - \Lambda}}{ R^2} \left [ 2 R^2
\Lambda,_z + \ e^{4\Psi} {\cal W} {\cal W},_z \right ] , 
\label{eq:emt4} \\
	&	&	\nonumber	\\
&S^2_2 &= \ \frac{e^{\Psi - \Lambda}}{ R^2} \left [ 2 R R,_z 
\right ] , \label{eq:emt5}
\end{eqnarray}\label{eq:semt}\end{subequations} 
and the nonzero components of the surface current density $\mbox{\sl j}_a$ are
\begin{subequations}
\begin{eqnarray}
& \mbox{\sl j}_t &= \ -2 e^{\Psi - \Lambda} A _{t,z} , 
\label{eq:corelec}   \\
&	&	\nonumber	\\
&\mbox{\sl j}_{\varphi} &= \ -2 e^{\Psi - \Lambda} A _{\varphi ,z}, 
\label{eq:cormag} 
\end{eqnarray}\label{eq:cor}\end{subequations}
where all the quantities are evalua\-ted at $z = 0^+$.

We now consider a solution with the above mentioned discontinuity properties by
taking the Taub-NUT-type solution (\ref{eq:tnt}) with $F(\nu)$ given by
\begin{equation}
F (\nu) \ = \ \nu + \alpha \sqrt{\nu^2 - 1} \ , \label{eq:fnu}
\end{equation}
where $\alpha \geq 0$. This choice for $F(\nu)$ was firstly presented in
reference  \cite{GL1} and leads to thin disks with  nonzero radial pressure and
of finite radius, located at  $z = 0$, $0 \leq r \leq 1$. In order to obtain
disks with  non-unit radius, we only need to make the transformation  $r
\rightarrow a r$, where $a$ is the radius of the disk.

With this choice for $F(\nu)$ the image of the disk by the  conformal mapping
(\ref{eq:conf}) is the surface 
\begin{equation}
\alpha ^2 R^2 + Z^2 = \alpha ^2
\end{equation}
so that the disks are mapped into spheroidal thin shells of  matter and its
exterior  is mapped into the exterior of   shells. We have three possible
values for $\alpha$:  $\alpha = 1$, a spherical shell, $\alpha >1$, a prolate 
spheroidal shell, and $0<\alpha<1$, an oblate spheroidal  shell. Therefore, 
the coordinates naturally adapted to the symmetry of the shells are,
respectively,  spherical, prolate spheroidal and oblate spheroidal.  By
considering the above shells as sources, we then seek for exterior solutions of
the  field equations (\ref{eq:ernst}) and, using (\ref{eq:comf}), we  obtain
the corresponding disk solutions in the original  coordinates.  Thus, for
example,  for the Taub-NUT-type  metric (\ref{eq:tnt}), written in prolate
spheroidal coordinates, we must choose  $\alpha >1$ and the shell would be 
located at $x =  \alpha / k > 1$ and $y= \sqrt {1-r^2}$, with $k= \sqrt {\alpha
^2 -1}$.
 
Due to the fact that the metric function ${\cal W} (r,z)$ depends only of the
prolate spheroidal coordinate $y$, and as a consequence of the behavior of the
derivatives of $F(\nu)$, we can easily see that ${\cal W},_z = 0$ at the disk
and so we have $S^1_0 = 0$ and $S^0_1 = (S^0_0 - S^1_1){\cal W}$. That is, the
SEMT can be written as an upper right triangular matrix and so can be easily
diagonalized in terms of an orthonormal tetrad ${{\rm e}_{\hat a}}^b = \{ V^b ,
W^b , X^b , Y^b \}$, where
\begin{subequations}\begin{eqnarray}
V^a &=& e^{- \Psi} \ ( 1, 0, 0, 0 )  ,	\\
	&	&	\nonumber	\\
W^a &=& \frac{e^\Psi}R \ \ ( -{\cal W}, 1, 0, 0 )  ,	\\
	&	&	\nonumber	\\
X^a &=& e^{\Psi - \Lambda} ( 0, 0, 1, 0 )  ,	\\
	&	&	\nonumber	\\
Y^a &=& e^{\Psi - \Lambda} ( 0, 0, 0, 1 ) .
\end{eqnarray}\label{eq:tetrad}\end{subequations}
Since the vectors ${{\rm e}_{\hat a}}^b$ are the eigenvectors of the SEMT, the
timelike vector $V^a$ can be interpreted as the velocity vector of the disks
and so the orthonormal tetrad is comoving. This is the orthonormal tetrad used
by static observers who are at rest with respect to infinity, or ``Locally
Static Observers'' (LSO) \cite{KBL}.

We obtain for the metric, the SEMT and the surface current density the
expressions
\begin{subequations}\begin{eqnarray}
g_{ab} \ &=& \ - V_a V_b + W_a W_b + X_a X_b + Y_a Y_b  ,
\label{eq:metdia}		\\
&   & \nonumber						\\
S_{ab} \ &=& \ \epsilon V_a V_b + p_\varphi W_a W_b +
p_r X_a X_b , \label{eq:emtdia} \\
&   & \nonumber						\\
\mbox{\sl j}_a \ &=& \ \sigma V_a + \mbox{\sl j} W_a, 
\label{eq:ja} 
\end{eqnarray}\end{subequations}
where
\begin{equation}
\epsilon \ = \ - S^0_0, \quad  p_\varphi \ = \ S^1_1, \quad 
 p_r \ = \ S^2_2,  \label{eq:dps}
\end{equation}
are, respectively, the surface energy density, the azimuthal  pressure, and the
radial pressure measured by this observer, and
\begin{equation}
\sigma = - V^0 \mbox{\sl j}_0,  \quad \mbox{\sl j}  =  W^0 \mbox{\sl j}_0 + W^1
\mbox{\sl j}_1,  \label{eq:djs}
\end{equation}
are the electric charge density and the azimutal current density  of the disk
measured by this observer.

From the expressions (\ref{eq:tnt}) for the Taub-NUT-type solution, is easy to
see that $p_\varphi = p_r = p$ and so the  surface energy-momentum tensor can
be cast in the perfect fluid form 
\begin{equation}
S_{ab} \ = \ (\epsilon + p)V_a V_b + ph_{ab},
\end{equation}
where $h_{ab} = g_{ab} - Y_a Y_b$ is the metric of the  $z=0$ hipersurface, 
\begin{subequations}
\begin{eqnarray}
\epsilon &=& \ - \left[\frac{\bar\alpha + a}{2k^2} \right] \left( \frac{p}{\bar
\alpha} \right)^3 , \label{eq:ener}  \\ 
p &=& \ \frac{2 \bar\alpha}{\sqrt{\bar\alpha^2 + 2a  \bar\alpha + 1 + c^2}}, 
\label{eq:pres}
\end{eqnarray} \label{eq:ep}\end{subequations}
where $\bar \alpha = \alpha /k > 1$. From (\ref{eq:ener})  follows that the
fluid has a  barotropic equation of state. The effective Newtonian density,
defined  as $\varrho=\epsilon +2p$, is   
\begin{equation}
\varrho= \left[ \frac{a (\bar \alpha ^2 + 1) + (2 + c^2) \bar\alpha}{2} \right]
\left( \frac{p}{\bar\alpha} \right)^3 . 
\end{equation}
The surface current density in the coordinates frame is given by
\begin{subequations}
\begin{eqnarray}
\mbox{\sl j}_t &=& c \left[ \frac{a \bar\alpha^2 + (1 + c^2)(2\bar\alpha +
a)}{8 \sqrt{2} k^2 (1 + c^2)^{1/2}} \right] \left( \frac{p}{\bar\alpha} 
\right)^5  ,    \\
\nonumber     \\
\mbox{\sl j}_\varphi &=& 2kb(1-r^2)^{1/2} \mbox{\sl j}_t   ,
\end{eqnarray}
\end{subequations}
whereas the surface electric charge density measured by  the comoving observer
is
\begin{equation}
\sigma = - \ c \left[ \frac{a \bar\alpha^2 + (1 + c^2)
(2 \bar\alpha + a)} {4 \sqrt{2} k(1+c^2)^{1/2}} \right] \left (
\frac {p}{\bar \alpha} \right ) ^4 , \label{eq:car}
\end{equation}
and the azimuthal current density measured by the comoving observer is equal to
zero. Thus we have a family of finite charged perfect fluid disks with constant
surface energy density and pressure given by (\ref{eq:ep}), and constant
surface  electric charge density given by (\ref{eq:car}).

\section{The physical properties of the disks}

We will now analyze the physical properties of the disks. In first instance we
consider that $c = 0$, so that $a^2 + b^2 = 1$. In this case the
electromagnetic potentials are equal to zero and so the disks are made of a
neutral perfect fluid. Furthermore, is easy to see that with $|a| \leq 1$ the
energy density is always a negative quantity and so the disks obtained when $c
= 0$ never satisfy the weak energy  condition. So, in order to have disks in
agreement with the weak energy condition, we need to consider solutions with $c
\neq 0$. That is, disks with nonzero surface electric charge density.

From expression (\ref{eq:ener}) is easy to see that the energy density is
positive if we take
\begin{equation}
a < - \bar \alpha < - 1.
\end{equation}
Additionally, in order to have real positive expressions for the pressure $p$
and the effective Newtonian density $\varrho$,  we must to impose the
conditions
\begin{subequations}\begin{eqnarray}
\bar \alpha ^2  + 2a \bar \alpha + 1 + c^2&>& 0, 	\\
&& \nonumber \\
a( \bar \alpha ^2 + 1) + (2+c^2) \bar \alpha &\geq& 0.
\end{eqnarray}\end{subequations}
Thus, for example, if we take  $c = 3$, $a = - 3$ and ${\bar \alpha} = 2$, we
have that
\begin{subequations}\begin{eqnarray}
\epsilon &=& 3 \sqrt{2}, 	\\
&& \nonumber \\
p &=& 2 \sqrt{2} ,  \\
&& \nonumber \\
\varrho &=& 7 \sqrt{2} , \\
&& \nonumber \\
\sigma &=& 0.6 \sqrt{15} ,
\end{eqnarray}\label{eq:exam}\end{subequations}
in agreement with the weak and strong energy conditions. Furthermore, $ p <
\epsilon$ and so the disk also satisfies the dominant energy condition.

In order to analyze the motion of the disks, we can see that the spatial
components of the velocity vector $V^a$ are zero with respect to the
coordinates and so the disks are ``locally static''. The motion of the disks
can also be analyzed by considering an orthonormal frame different from the
comoving tetrad. So, an orthonormal tetrad commonly used is the ``Locally 
Nonrotating Frame'' or ``Zero Angular Momentum Observer'' \cite{BPT,CHAN}. The
tangential velocity of the disks measured by this observer is given by  
\begin{equation}
v_{_{LNRF}} = \frac{g_{11}(\Omega - \omega)}{\sqrt {g_{01}^2 -  g_{00}g_{11}}},
\end{equation}
where $\omega = - g_{01}/g_{11}$ and $\Omega = V^1/V^0$. We obtain
\begin{equation}
v_{_{LNRF}} \ = \ - \frac b {2k} \left ( \frac p {\bar \alpha}  \right ) ^2 
\frac{\sqrt{1-r^2}}{r}.
\end{equation}
As we can see from the above expression, with respect to LNRF, the particles of
the disks move with superluminal velocities for $r < r_0$, where
\begin{equation}
r_0 = \frac{b p^2}{\sqrt{b^2p^4 + 4k^2 \bar \alpha ^4}}.
\end{equation}
For instance, with the values used in (\ref{eq:exam}), we have that $r_0 =
\sqrt{12/13} \approx 0.96$, and so the disks will have subluminal velocities
only in a narrow region near the edge. However, the above superluminal behavior
of the velocity is indeed due to the motion of the LNRF and not to the motion
of the disks. So, we can see that for this family of solutions the LNRF is not
a well behaved observer in the sense that the frame itself presents
superluminal velocities. 

Another quantity related with the  motion of the disk is  the specific angular
momentum of a particle rotating at a  radius $r$, defined as $h =  g_{\varphi
a}V^a$. Thus we have 
\begin{equation}
h^2 = \left (\frac{bp}{\bar \alpha}\right ) ^2 \sqrt{1-r^2}. 
\end{equation}
This quantity can be used to analyze the stability of the  disks against radial
perturbations. The condition of stability,
\begin{equation}
\frac{d(h^2)}{dr} \ > \ 0  ,
\end{equation}
is an extension of Rayleigh criteria of stability of a fluid  in rest in a
gravitational field \cite{FLU}. As we can see, the disks are not stable under
radial perturbations.

\section{discussion}

The interpretation of a family of electrovacuum stationary Taub-NUT-type 
fields in terms of charged perfect fluid disks of finite extension  was
presented. The interpretation was done by means of an ``inverse problem"
approach used to obtain disk sources of known solutions of the Einstein or
Einstein-Maxwell equations.  The diagonalization of the energy-momentum tensor
of the disks was facilitated by the fact that  it  was written as an upper
right triangular matrix.  We find that the  inclusion of electromagnetic fields
changes significatively the different material properties of the  disks and so,
for some values of the parameters, finite charged perfect fluid disks were
obtained that are in agreement with all the energy conditions. However, the
disks are not stable under radial perturbations. We also find that the disks
have  a  barotropic equation of state.

\begin{acknowledgments}

The authors want to thank the finantial support  from  COLCIENCIAS, Colombia.

\end{acknowledgments}

\end{document}